\PassOptionsToPackage{usenames,dvipsnames}{xcolour}
\documentclass[twocolumn,twocolappendix,nofootinbib,iop]{openjournal}

\pdfoutput=1 
\usepackage{amsmath,amssymb,amstext}
\usepackage[T1]{fontenc}
\usepackage{apjfonts}
\usepackage{ae,aecompl}
\usepackage{tikz}
\usetikzlibrary{arrows.meta, positioning}
\usepackage[utf8]{inputenc}
\usepackage[colorlinks,allcolors=blue]{hyperref}
\usepackage[figure,figure*]{hypcap}
\usepackage{natbib}
\usepackage{url}
\usepackage{mdwlist}
\usepackage{multirow}
\usepackage{lineno}
\usepackage{fontawesome}
\usepackage{cleveref}

\usepackage{seqsplit}

\newcommand{\hash}[1]{{\ttfamily\seqsplit{#1}}}

\crefname{figure}{Fig.}{Figs.}  
\Crefname{figure}{Figure}{Figures}  

\usepackage{soul}

\usepackage{xcolor}
\definecolor{maroon}{cmyk}{0, 0.87, 0.68, 0.32}
\definecolor{halfgray}{gray}{0.55}
\definecolor{ipython_frame}{RGB}{207, 207, 207}
\definecolor{ipython_bg}{RGB}{247, 247, 247}
\definecolor{ipython_red}{RGB}{186, 33, 33}
\definecolor{ipython_green}{RGB}{0, 128, 0}
\definecolor{ipython_cyan}{RGB}{64, 128, 128}
\definecolor{ipython_purple}{RGB}{170, 34, 255}

\usetikzlibrary{shapes.geometric, arrows}

\tikzstyle{process} = [rectangle, rounded corners, minimum width=3cm, minimum height=1cm, text centered, draw=black, fill=blue!20]
\tikzstyle{data} = [parallelogram, minimum width=3cm, minimum height=1cm, text centered, draw=black, fill=green!20]
\tikzstyle{decision} = [diamond, minimum width=3cm, minimum height=1cm, text centered, draw=black, fill=red!20]
\tikzstyle{arrow} = [thick,->,>=stealth]

\usepackage{listings}

\newcommand{\githubtopo}{\href{https://github.com/ChristianFidler/Topo/}{\faGithub}}

\newcommand{\githubcobaya}{\href{https://github.com/CobayaSampler/cobaya}{\faGithub}}
\newcommand{\githubtopocobaya}{\href{https://github.com/santiagocasas/topo-cobaya}{\faGithub}}

\begin{document}
\journalinfo{The Open Journal of Astrophysics}

\title{\texttt{TOPO}: Time-Ordered Provable Outputs}


\author{
Santiago Casas$^{1, 2,\ast}$, 
Christian Fidler$^{1, \dagger}$
}
\thanks{$^\ast$ santiago.casas@port.ac.uk}
\thanks{$^\dagger$ fidler@physik.rwth-aachen.de}

\affiliation{
$^1$ Institute for Theoretical Particle Physics and Cosmology (TTK), RWTH Aachen University, 52056 Aachen, Germany
}
\affiliation{$^2$ Institute of Cosmology and Gravitation, University of Portsmouth, Dennis Sciama Building,
Portsmouth, PO1 3FX, UK}


\begin{abstract}
We present \texttt{TOPO} (Time-Ordered Provable Outputs) \githubtopo, a tool designed to enhance reproducibility and data integrity in astrophysical research, providing a trustless alternative to data analysis blinding. Astrophysical research frequently involves probabilistic algorithms, high computational demands, and stringent data privacy requirements, making it difficult to guarantee the integrity of results. \texttt{TOPO} provides a secure framework for verifying reproducible data analysis while ensuring sensitive information remains hidden. Our approach utilizes deterministic hashing to generate unique digital fingerprints of outputs, and Merkle Trees to store outputs in a time-ordered manner. This enables efficient verification of specific components or the entire dataset while making it computationally infeasible to manipulate results—thereby mitigating the risk of human interference and confirmation bias, key objectives of blinding methods.
We demonstrate \texttt{TOPO}'s utility in a cosmological context using \texttt{TOPO-Cobaya} \githubtopocobaya, showcasing how blinding and verification can be seamlessly integrated into Markov Chain Monte Carlo calculations, rendering the process cryptographically provable. This method addresses pressing challenges in the reproducibility crisis and offers a robust, verifiable framework for astrophysical data analysis.

\end{abstract}

\maketitle


\section{Introduction}

Astrophysics, a field built on precise data analysis and the ideal of reproducibility, faces unique challenges in safeguarding the integrity and confidentiality of its findings. One of the most critical aspects is mitigating human confirmation bias—a well-documented concern in scientific analysis \cite{KlaymanHa1987, MacCounPerlmutter2015, 2011arXiv1112.3108C}. To address this, blinding techniques have become essential tools, ensuring that results remain objective and unbiased by expectations or preconceptions.
Current blinding methods in astrophysics include:

\begin{enumerate}
    \item \textbf{Altering Data}:
    One classical example in cosmology comes from \cite{Davis_2007}, where supernova data was subtly altered by stretching redshifts. Another common approach involves modifying cosmological parameters entering the equations of the analysis, effectively "blinding" the final results \cite{Zhang10.1093/mnras/stx1600}. However, this method has a drawback—once the data is unblinded, analyses often need to be re-run, especially to recalibrate for unknown systematic errors.

    \item \textbf{Catalog-level blinding}: In this method, entire data catalogs are systematically modified, affecting cosmological parameter estimates. Specific parameters—such as object positions, ellipticities, magnitudes, or redshifts—are shifted in a controlled manner \cite{Brieden_2020, Muir2019Blinding}. This technique has been used by the DESI collaboration \cite{DESI:2024cdy}, the DES collaboration  \cite{Asgari}, and the KIDS collaboration \cite{KIDS1000}, as originally proposed by \cite{10.1093/mnras/stv2140}. Researchers analyze these altered catalogs without knowing the exact modifications, allowing for unbiased analysis and systematic validation. However, this method relies on an external “unblinder”—a trusted individual who reveals the key to the transformations before the final results are published, raising concerns about trust and timing.
    
    \item \textbf{Shifting Covariances}: Proposed by \cite{sellentin2020blinding}, this technique involves altering data covariance matrices, central to cosmological parameter estimation, in a reversible manner. By shifting the covariance matrix, the best-fit point from a Bayesian likelihood analysis is shifted as well, allowing the blinder to hide the "true" parameter estimation. This approach avoids directly manipulating data points, making accidental unblinding more difficult. However, like other methods, it still depends on a trusted individual and is limited to certain types of cosmological likelihood functions.
\end{enumerate}

While current blinding methods have proven effective, they are not without limitations. Existing approaches, such as data alteration, catalog-level blinding, and covariance matrix shifting, still rely heavily on trust. This creates potential vulnerabilities, as blinding often depends on a single individual or external entity—referred to as the unblinder—who controls the key to the modifications. This reliance introduces a risk of bias, either through accidental exposure or potential manipulation, which could compromise the scientific integrity of the results. Also the process of unblinding is irreversible meaning that after the analysis is unblinded it is not possible to prove any statements about the integrity of following analysis on the data.

To address these challenges, we introduce \texttt{TOPO} (Time-Ordered Provable Outputs) \footnote{\texttt{https://github.com/ChristianFidler/Topo/}}, a novel tool designed to provide a fully trustless and cryptographically secure method for conducting blinded analyses. \texttt{TOPO} goes beyond traditional methods by implementing a framework that enables the verification of a pipeline without exposing sensitive information. This ensures that results can be validated without the need to rely on any individual, significantly reducing the risk of human error or interference.
Our blinding method \texttt{TOPO} can be used in addition to any blinding implemented by the survey and can be adopted on an individual basis as an additional way to create mathematical trust on the results, at almost no additional cost to the user.  

The key innovation of \texttt{TOPO} lies in its use of \textit{deterministic hashing} and \textit{Merkle Trees}. Deterministic hashing generates unique digital fingerprints of outputs, ensuring that each result is reproducible and verifiable. These digital fingerprints provide a robust, tamper-proof method of confirming the authenticity and integrity of the data. 

Hashes are widely used across various fields beyond astrophysics, playing a critical role in ensuring data integrity and security, see \cite{sobti2012cryptographic} for a review. In the world of banking and financial services, hash functions are fundamental to safeguarding sensitive data such as transaction records and digital signatures, enabling secure online payments and preventing fraud. Similarly, in internet security, hashes are employed to verify the integrity of files during downloads or updates, ensuring that the transmitted data has not been altered. Cryptographic hashes are also essential in generating public-private key pairs in encryption protocols, which are the backbone of secure communications across the web. This widespread reliance on hash functions underscores their versatility and importance in creating trustless systems where data integrity is paramount.

\texttt{TOPO} further leverages Merkle Trees, an efficient cryptographic primitive, designed for structured data. They allow for an efficient verification of both individual components and the entire dataset, maintaining computational efficiency. Crucially, organizing outputs in this way makes it computationally infeasible to manipulate or alter results afterwards, addressing a key weakness in traditional blinding methods. This framework not only mitigates confirmation bias but also enhances the transparency and reproducibility of astrophysical research.

In contrast to traditional methods, \texttt{TOPO} offers a \textit{fully trustless system}, meaning that no individual or entity holds exclusive control over the verification process. Through the integration of cryptographic proofs, \texttt{TOPO} guarantees that a specific pipeline and dataset were used to produce the results, eliminating the need to rely on trust in the unblinding process. This represents a significant step forward in addressing the reproducibility crisis that has plagued academic research in recent years.

We demonstrate \texttt{TOPO}’s practical application in a cosmological context using \texttt{TOPO-Cobaya}\footnote{\texttt{https://github.com/santiagocasas/topo-cobaya}}, a code that integrates \texttt{TOPO}’s blinding and verification framework into Markov Chain Monte Carlo (MCMC) calculations, forking the popular likelihood sampler code \texttt{Cobaya} \cite{Torrado:2020dgo} \githubcobaya. \texttt{TOPO-Cobaya} allows researchers to carry out cosmological parameter estimation while maintaining the highest standards of data integrity, providing cryptographic proof of the results at every stage of the analysis. This approach ensures that the entire analysis pipeline is not only reproducible but also verifiable in a manner that is transparent and resistant to manipulation.

By employing \texttt{TOPO} in astrophysical research, we aim to address some of the most pressing challenges related to reproducibility and human bias in data analysis. As the complexity of cosmological data grows with large-scale surveys and high-performance computational methods, the need for trustless, verifiable analysis pipelines becomes increasingly critical. \texttt{TOPO} offers a pathway to achieve this, providing the field with a robust, transparent framework for blinded analysis that ensures the integrity and confidentiality of scientific findings.

\definecolor{codebg}{rgb}{0.95,0.95,0.95}
\lstset{
    backgroundcolor=\color{codebg},
    basicstyle=\ttfamily\footnotesize,
    keywordstyle=\color{blue},
    stringstyle=\color{green},
    commentstyle=\color{gray},
    showstringspaces=false,
    frame=single,
    breaklines=true,
    captionpos=b
}

\section{Cryptographic Basics}
In this section, we briefly outline the key cryptographic concepts used throughout this paper.

\subsection{Hashes}

A hashing function \( f \) is a mathematical trapdoor function with the following properties:

\begin{enumerate}
    \item It is fast to compute.
    \item It is computationally infeasible to reverse.
    \item Small differences in the input result in vastly different outputs. The output is unique to each input, i.e., hash collisions are extremely rare or nonexistent.
    \item The output, or hash, has a fixed length regardless of the size of the input.
\end{enumerate}

We employ the widely-used \texttt{SHA256} hashing function. The \texttt{SHA256} hash of the last point is \texttt{16ec...\allowbreak06a9}, displaying only the first and last four characters for brevity. To give an example of point 3, if you replace "length" by "lenght" in the same sentence, you get: \texttt{e2d2...\allowbreak93d4}. As we can see, a small change in the input changes completely the output and it is impossible to infer the input from any number of outputs\footnote{At \texttt{https://www.movable-type.co.uk/scripts/sha256.html} you can experiment and see the SHA256 output for any given input and a code implementation}.

Utilizing these properties, a hash \( H = f(P) \) serves as a digital fingerprint for any data \( P \) (also known as the pre-image). The hash is a short, unique representation of the data that doesn’t reveal any details about the data itself. However, it can be used to quickly verify if the data has been changed or is still intact.

For example, \texttt{Git} \footnote{See the hash object documentation \texttt{https://git-scm.com/docs/git-hash-object}} employs this technique by generating a unique hash for each commit. After downloading the code, the integrity of the code can be verified by comparing the locally computed hash with the original commit hash. The chain of commit hashes, makes sure that the version history of the code has been preserved.

\subsection{Merkle Trees}

A Merkle tree \cite{merkle1987digital} is a fundamental data structure in cryptography and computer science, particularly relevant in distributed systems, blockchain technologies, and secure communication protocols. It is designed to efficiently and securely verify the integrity of large datasets by leveraging cryptographic hashes. The structure of a Merkle tree allows for compact and scalable data verification, making it ideal for systems where bandwidth, storage, and computational efficiency are key concerns.

\subsubsection{Structure and Function}

\begin{figure}[h]
\centering
\begin{tikzpicture}[
    level distance=2.cm,
    level 1/.style={sibling distance=40mm},
    level 2/.style={sibling distance=20mm},
    level 3/.style={sibling distance=20mm},
    every node/.style={circle, draw, fill=blue!20, minimum size=0.4cm},
    red node/.style={fill=red!50} 
]
    \node [red node] {Merkle Root}
        child { node[red node] {Hash $1 \| 2$}
            child { node[red node] {Hash 1} 
                child {node {Data 1}}}
            child { node {Hash 2} 
                child {node {Data 2}}}
        }
        child { node {Hash $3 \| 4$}
            child { node {Hash 3}  
                child {node {Data 3}}}
            child { node {Hash 4}       
                child {node {Data 4}}}
        };

    \draw[->, thick] (-3,-7) -- (3,-7);
    
    \draw (0,-7.2) node[draw=none,fill=none] {\textit{Time}};
    
\end{tikzpicture}
\caption{A representation of a Merkle tree, where each node contains the hash of its child nodes. The Merkle root is derived from all leaf data, ensuring data integrity and efficient verification. The leaves represent individual data blocks, while the internal nodes are their respective hashes. In the left branch of the tree (shown in red), the colored nodes represent the proof object submitted for verification in \texttt{TOPO}. The arrow at the bottom signifies the time-ordered structure of the data.}
\label{fig:angela}
\end{figure}
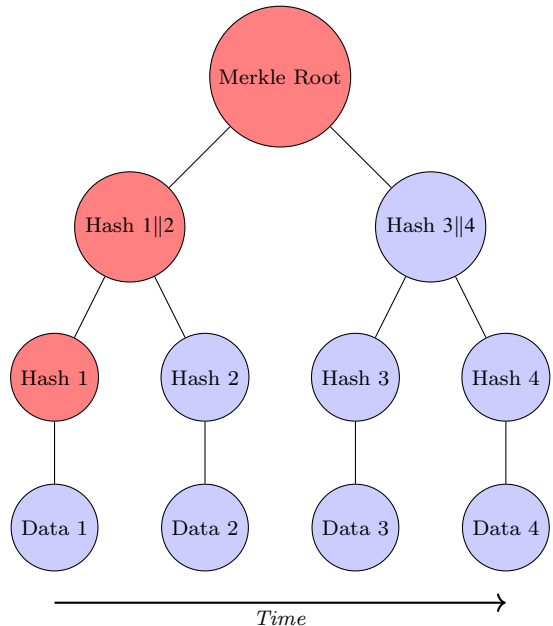

A Merkle tree is constructed by hashing individual data blocks \( L_n \), which form the leaves of the tree. Each data block is hashed individually: \( H_n = H(L_n) \). Adjacent pairs of these hashed values are then concatenated (the $\|$ operator) and hashed again to form the next level of the tree:
\[
H_{ab} = H(H_a \| H_b)
\]
This process is repeated, layer by layer, until a single hash remains, called the \textbf{Merkle root} (see \cref{fig:angela}). The Merkle root serves as a unique fingerprint for the entire dataset, ensuring that any modification to even a single data block results in a different root hash. The tree structure scales logarithmically, meaning that the number of layers grows slowly with the number of data blocks, making it highly efficient for large datasets.

\subsubsection{Example Construction}

\begin{enumerate}
    \item \textbf{Leaf Nodes}: The leaves of the tree are the hashes of individual data blocks. For example, given four data blocks \(D_1, D_2, D_3, D_4\), the corresponding hashes are \(H(D_1)\), \(H(D_2)\), \(H(D_3)\), and \(H(D_4)\).
    
    \item \textbf{Intermediate Nodes}: The next level is created by concatenating and hashing pairs of leaf nodes:
    \[
    H_{1\|2} = H(H(D_1) \| H(D_2))
    \]
    \[
    H_{3\|4} = H(H(D_3) \| H(D_4))
    \]
    
    \item \textbf{Root Node}: The Merkle root is computed by hashing the top-level concatenation of the intermediate nodes:
    \[
    H_{\text{root}} = H(H_{1\|2} \| H_{3\|4})
    \]
\end{enumerate}

\subsubsection{Applications and Advantages}

Merkle trees provide numerous benefits in the verification and integrity of data:

\begin{itemize}

    \item \textbf{Tamper Resistance and Integrity:} Similar to a simple hash, any change to a single data element alters the entire Merkle root, making even the smallest modifications immediately detectable. The Merkle root, a fixed-length hash, serves as a summary of the entire dataset. This compact representation enables efficient storage and transmission of integrity information, ensuring both data integrity and tamper resistance.
    
    \item \textbf{Scalability:} As the number of data blocks $N$ increases, the number of layers of the Merkle tree grows logarithmically as $\mathcal{O}(\log_2 (N))$. For instance, for 4 data blocks, the tree will have 3 layers, where the $+1$ comes from the final Merkle root. This makes it an ideal structure for large-scale datasets, where efficient computation and storage are critical.   
    
    \item \textbf{Proof of Inclusion:} Merkle trees allow for the creation of efficient proofs that a specific data block is part of the dataset (proof of inclusion). These proofs can be generated and verified with minimal computational overhead, thanks to the logarithmic scaling.

\end{itemize}

\subsection{Elliptic Curve Digital Signature Algorithm (ECDSA)}

The Elliptic Curve Digital Signature Algorithm (ECDSA), detailed in \cite{ANSI-X9.62-1999}, is a cryptographic algorithm used to generate and verify digital signatures. It is based on elliptic curve cryptography (ECC), see \cite{cryptoeprint:2024/1265} for a review. ECDSA offers a high level of security with small key sizes, compared to algorithms like RSA (see \cite{Rivest1978}), which are widely use on the internet, for instance in the HTTPS protocol and are based on prime number factorization.

\subsubsection{Key Pair Generation}
ECDSA requires two keys:
\begin{itemize}
    \item \textbf{Private Key}: A randomly generated number that must be kept secret. It is used for signing messages.
    \item \textbf{Public Key}: A unique identifier derived from the private key through elliptic curve mathematics. It is used to verify signatures and can be shared publicly.
\end{itemize}

\subsubsection{Properties of ECDSA}
For our purposes, the key advantage of ECDSA is its ability to create tamper-proof digital signatures, ensuring that only authorized individuals, with acces to the private key, can sign the proof object in TOPO, confirming the calculation's authenticity. Additionally, the signature guarantees that the signed data has not been altered.

\section{Blinding Analysis with \texttt{TOPO}}

In this section, we outline how hashes, Merkle Trees, and ECDSA keys are used to design a general blinding procedure applicable to any pipeline in astrophysics. Our goal is to create a trustless process—a system in which verification does not require relying on a trusted third party. Instead, cryptographic methods ensure that the integrity of the analysis can be independently confirmed by anyone.

Using these cryptographic tools, we provide a public and transparent approach to analysis verification, allowing any member of the scientific community to independently verify the claims without the need to trust any specific individual or organization. This process leverages deterministic hashing to generate unique digital fingerprints, Merkle Trees to organize outputs efficiently, and ECDSA keys to authenticate the results, ensuring that the analysis is secure and tamper-proof. We name this approach \texttt{TOPO}: Time-Ordered Provable Outputs.

In the following we define the individual conducting the analysis as the \textit{Prover} and the one verifying the \textit{Prover}'s honesty as the \textit{Verifier}.

\subsection{Freezing the Analysis}
\begin{figure}
\begin{center}
\begin{tikzpicture}[
    node distance=1.2cm, auto,
    process/.style={rectangle, draw, fill=blue!20, text width=5cm, align=center, minimum height=1cm},
    arrow/.style={thick,->,>=stealth}
]
    \node (code) [process] {Git commit hashes of codes $H_{\text{codes}}$};
    \node (input) [process, below of=code] {SHA256 hash of input files $H_{\text{input}}$};
    \node (combine) [process, below of=input] {Combine $H_{\text{codes}} \| H_{\text{input}}$};
    \node (analysis) [process, below of=combine] {Produce Analysis-Hash \\ $A = \text{SHA256}(H_{\text{codes}} \| H_{\text{input}})$};
    \node (blockchain) [process, below of=analysis] {Sign $A$ as $S_A$ using the private key associated with the public key $P$};
    \node (sign) [process, below of=blockchain] {Release commitment object $C=\{A, S_A, P\}$};
    \node (frozen) [process, fill=green!20, below of=sign] {Analysis pipeline is now frozen};

    \draw[arrow] (code) -- (input);
    \draw[arrow] (input) -- (combine);
    \draw[arrow] (combine) -- (analysis);
    \draw[arrow] (analysis) -- (blockchain);
    \draw[arrow] (blockchain) -- (sign);
    \draw[arrow] (sign) -- (frozen);
\end{tikzpicture}
\end{center}
\caption{A flowchart illustrating the creation of the Analysis-Hash. The Prover starts by collecting the Git commit hash of the codes and the hash of the input files. These hashes are combined to generate a single Analysis-Hash, which serves as a verifiable identifier for the entire analysis pipeline.}
\label{fig:commit}
\end{figure}
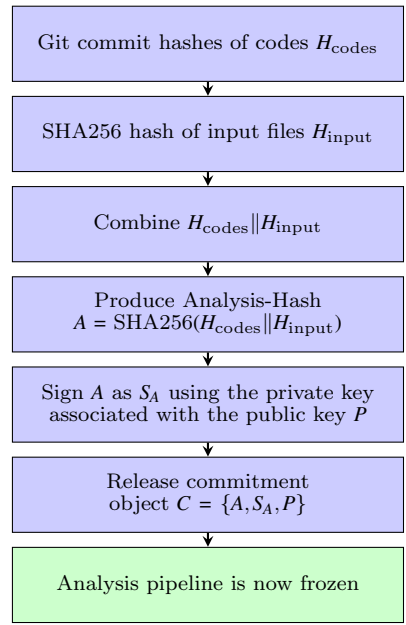

Initially, the \textit{Prover} finalizes their analysis pipeline. At this point, they create a cryptographic commitment that summarizes the entire workflow. To achieve this, we employ cryptographic hashes. This commitment flow is visualized in \cref{fig:commit}.
The Prover collects the Git-commit hashes of their employed codes along with a hash of the intended input files. These input files can include configuration files, numerical constants and other settings needed for the analysis code. These individual hashes are then combined into a single hash, producing the so-called Analysis-Hash:

\[
A = \textit{SHA256}(H_{\rm codes} || H_{\rm input}),
\]

which serves as a verifiable identifier of the analysis pipeline.

The Analysis-Hash must be stored on a public trusted medium that provides timestamping. This process locks in the analysis pipeline by allowing any Verifier to reconstruct the hash from a given pipeline and confirm that it matches the version committed by the Prover at the timestamp, ensuring no changes have been made since.

A possible solution for storing this Analysis-Hash is via platforms like the \texttt{arXiv} \footnote{\texttt{https://arxiv.org/}}, where anyone can download a local copy and verify the data's integrity. However, a more robust solution is to store the Analysis-Hash on a public blockchain, such as Ethereum \footnote{\texttt{https://ethereum.org/}}, which ensures both trust, irreversibility and time-stamping.

Finally, the Prover must verify their identity. To achieve this, we employ cryptographic signatures. The Prover publishes a unique public key tied to their identity (for example, in a published paper) and signs the Analysis-Hash with that key. Both the Analysis-Hash and its signature should then be published together.

\begin{figure}[t] 
\label{fig:prover}
\begin{center}
\begin{tikzpicture}[
    node distance=1.2cm, auto,
    process/.style={rectangle, draw, fill=blue!20, text width=5cm, align=center, minimum height=1cm},
    arrow/.style={thick,->,>=stealth}
]
    \node (frozen) [process, fill=green!20, below of=sign] {Start from frozen analysis pipeline};
    \node (analysis) [process, below of=frozen, yshift=-0.5cm] {Perform analysis on full data};
    \node (result) [process, below of=analysis] {Create cryptographic proof of analysis using Merkle tree};
    \node (mainhash) [process, below of=result] {Publish cryptographic proof-object $M$, provide signature $S_M$};
    \node (frozen2) [process, fill=green!20, below of=mainhash] {Analysis is now ready for independent verification};

    \draw[arrow] (frozen) -- node[anchor=west] {\textbf{After data availability}} (analysis);
    \draw[arrow] (analysis) -- (result);
    \draw[arrow] (result) -- (mainhash);
    \draw[arrow] (mainhash) -- (frozen2);
\end{tikzpicture}
\end{center}
\caption{This flowchart outlines the steps taken by the Prover to ensure cryptographic verification of the analysis. The process begins with the preparation of the analysis pipeline, followed by generating fingerprints of the codes and input data to compute the Analysis-Hash. The Analysis-Hash is then published on the blockchain with a digital signature, and the public key is released. After freezing the pipeline, the Prover performs the analysis on the full dataset and creates a cryptographic proof using a Merkle tree. Finally, the proof, analysis pipeline, and input files are published, ensuring that the analysis is ready for independent verification.}
\end{figure}
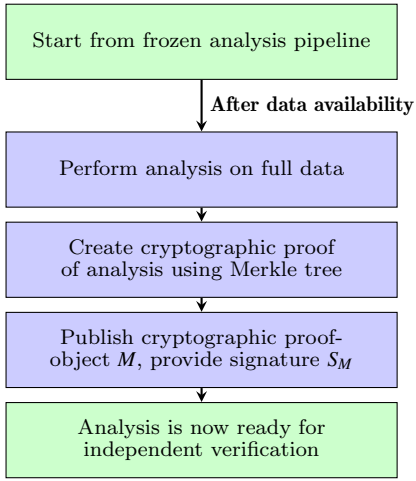

The signature serves two purposes: it prevents impostor attacks and links the frozen analysis to a known identity, stopping individuals from conducting numerous analyses and selectively publishing favorable ones. If the data is released but no analysis is published, the publicly identified Prover can be held accountable. Additionally, the signature ensures the authenticity of the analysis-hash and verifies that it has not been modified.

The procedure of the Prover is summarised in \cref{fig:commit} and consists of 
\begin{itemize}
    \item Freeze analysis pipeline consisting of code and input files.
    \item Collect git commit hashes of codes: $H_{\rm codes}$,
    \item Hash input files into $H_{\rm input}$,
    \item Produce Analysis-Hash \( A = \texttt{SHA256}(H_{\rm codes} || H_{\rm input}) \),
    \item Sign $A$ using private key associated to public key $P$ and produce $S_A$,
    \item Publish the commitment object $C$, composed of $C=\{A, S_A, P\}$ to a time-stamp server.
\end{itemize}

\subsection{Generating proof of honest analysis}

The next step is to analyze the actual data once it becomes available. The Prover then applies the frozen analysis pipeline to the data to obtain the results. After completing this step, the Prover generates and publishes a cryptographic proof demonstrating that the analysis has been performed as originally committed.

Proving that code has been executed as committed is a challenging problem, particularly for general-purpose code. Our approach requires two key properties of the analysis code: first, it must be deterministic, and second, it must produce a meaningful summary of its status during runtime. In this case, we can organize the summary output in a time-ordered manner, divide it into individual blocks, and combine them into a Merkle Tree.

By traversing the left-most branches of the Merkle tree, represented by the red nodes in figure \cref{fig:angela}, we observe that the first root depends only on the first data block, the second root depends on the first two blocks, the third on the first four blocks, and so on. Each higher root encompasses an exponentially larger portion of the data, with the topmost root serving as a unique identifier for the entire dataset.

This left-most branch of the Merkle tree enables verification of the computation at varying levels of computational cost. Lower levels can be verified quickly, while the topmost root requires a full recomputation of the code. Despite its complexity, the left-most branch remains a very compact structure, even for huge datasets.

The Prover now:
\begin{itemize}
    \item Publishes the left-most branch of the Merkle tree \( L \),
    \item Publishes the hash of the dataset used \( D \),
    \item Publishes the main hash \( H = \texttt{SHA256}(L || D) \),
    \item Publishes the signature \( S \) of \( H \),
    \item Makes the code and input files publicly available.
\end{itemize}

The entire procedure for the prover is summarized in \cref{fig:prover}.

\subsection{Verification Process}

With this information, anyone can verify the results and confirm the honest behavior of the Prover.

A Verifier first checks the signatures of the Analysis-Hash \( A \) and the main Hash \( H \) against the known public key of the prover. If the information is authenticated, they download the code and reconstruct the analysis pipeline, verifying that it matches the Analysis-Hash \( A \). At this point, the Verifier is assured that they possess the same analysis that the Prover committed to in the past proven by the timestamp on \( A \).

Next, the Verifier validates the data hash \( D \) against the dataset and begins executing the code. They can allocate as little or as much computational power as desired, computing the first roots of the Merkle tree. If these match the Prover's Merkle tree \( L \), the Verifier can be confident that they are running the exact same analysis as the Prover up to that point, and in principle, could recompute the entire analysis.

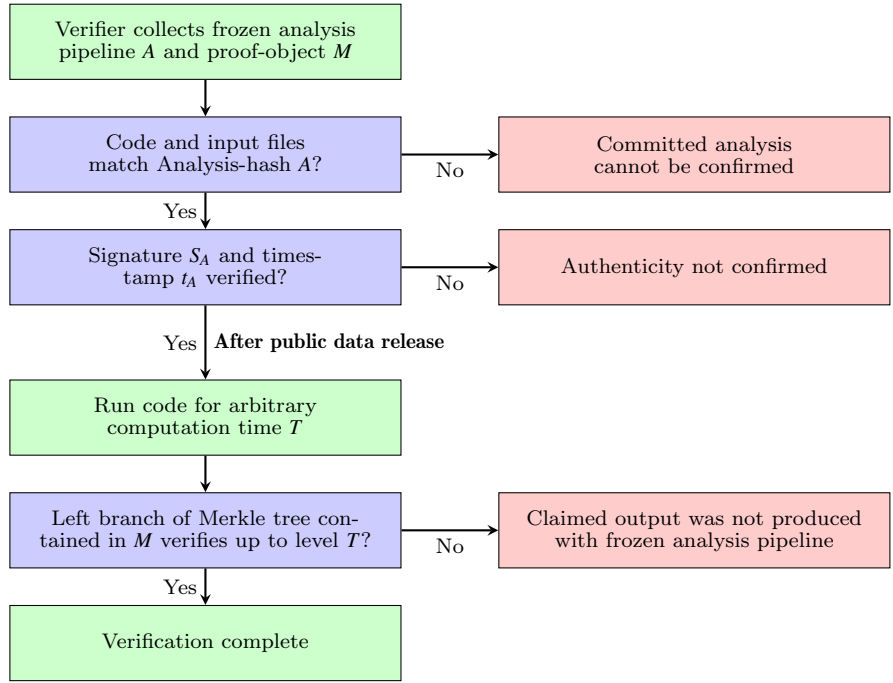
\begin{figure*}[htbp] 
\label{fig:verifier}
\begin{center}
\begin{tikzpicture}[
    node distance=1.5cm, auto,
    process/.style={rectangle, draw, fill=blue!20, text width=5cm, align=center, minimum height=1cm},
    decision/.style={rectangle, draw, fill=red!20, text width=5cm, align=center, minimum height=1cm},
    result/.style={rectangle, draw, fill=green!20, text width=5cm, align=center, minimum height=1cm},
    arrow/.style={thick,->,>=stealth}
]
    \node (verifier) [result] {Verifier collects frozen analysis pipeline $A$ and proof-object $M$};
    \node (verifycode) [decision, fill=blue!20, below of=verifier] {Code and input files match Analysis-hash $A$?};
    \node (fail1) [decision, right of=verifycode, xshift=5cm] {Committed analysis cannot be confirmed};
    \node (verifysign) [decision, fill=blue!20, below of=verifycode] {Signature $S_A$ and timestamp $t_A$ verified?};
    \node (fail2) [decision, right of=verifysign, xshift=5cm] {Authenticity not confirmed};
    \node (runcode) [result, fill=green!20, below of=verifysign, yshift=-0.5cm] {Run code for arbitrary computation time $T$};
    \node (verify) [decision, fill=blue!20, below of=runcode] {Left branch of Merkle tree contained in $M$ verifies up to level $T$?};
    \node (fail3) [decision, right of=verify, xshift=5cm] {Claimed output was not produced with frozen analysis pipeline};
    \node (end) [process, fill=green!20, below of=verify] {Verification complete};

    \draw[arrow] (verifier) -- (verifycode);
    \draw[arrow] (verifycode) -- node[anchor=east] {Yes} (verifysign);
    \draw[arrow] (verifysign) -- node[anchor=east] {Yes} node[anchor=west] {\textbf{After public data release}} (runcode);
    \draw[arrow] (runcode) -- (verify);
    \draw[arrow] (verify) -- node[anchor=east] {Yes} (end);
    \draw[arrow] (verifycode.east) -- node[anchor=north] {No} (fail1);
    \draw[arrow] (verifysign.east) -- node[anchor=north] {No} (fail2);
    \draw[arrow] (verify.east) -- node[anchor=north] {No} (fail3);
\end{tikzpicture}
\end{center}
\caption{This flowchart illustrates the process followed by the verifier to ensure the integrity of the analysis. The verifier begins by collecting the relevant components (code, input files and data). The verification steps include checking if the code and input match the committed Analysis-Hash, verifying the digital signatures, running the code, and validating the left branch of the Merkle tree. Any failure at each stage results in a failure to confirm the committed analysis, whereas passing all steps leads to a successful verification.}
\end{figure*}

This setup allows for full verification of the analysis while also providing the option for a quick test. The Verifier does not require any form of authentication and can be any member of the public with sufficient computational resources. Since the Prover has no knowledge of how much of the code will be verified, they are strongly incentivised to perform the calculations honestly. Especially since they signed the analysis cryptographically and any cheating could be detected and linked to their identity. This scheme encourages honest behavior while maintaining strict security.
The workflow of a verifier is summarized in figure \cref{fig:verifier}.

\section{\texttt{TOPO-Cobaya} : A CLI Tool for Verifiable Cosmological Analysis}
As a proof of concept, we have developed a version of \texttt{TOPO} compatible with \texttt{Cobaya} \cite{Torrado:2020dgo} for the computation of MCMC chains.
\texttt{TOPO-Cobaya} is a command-line interface (CLI) tool designed to facilitate verifiable cosmological analysis. It integrates seamlessly with the \texttt{Cobaya} framework, providing a suite of commands for key generation, analysis freezing, proof generation, and verification.

\subsection{Command Structure}

\texttt{TOPO-Cobaya} is invoked using the \texttt{topocobaya} command, followed by one of four subcommands:

\begin{enumerate}
    \item \texttt{keygen}: Generates or loads cryptographic keys
    \item \texttt{freeze}: Freezes the analysis state
    \item \texttt{proof}: Generates proof of the analysis
    \item \texttt{verify}: Verifies the analysis
\end{enumerate}

Each subcommand accepts specific arguments and options, providing a flexible and user-friendly interface.

\subsection{Key Generation}

The \texttt{keygen} command handles the generation and management of public/private key pairs used for signing. Users can either generate a new key pair or load an existing private key:

\begin{verbatim}
topocobaya keygen [--load KEY_PATH]
\end{verbatim}

If no existing key is specified, the tool prompts for confirmation before generating a new key pair.

\subsection{Freezing the Analysis}

The \texttt{freeze} command is used to capture the initial state of the analysis:

\begin{verbatim}
topocobaya freeze INPUT.yaml [-p INPUT.yaml]
\end{verbatim}

This command:
\begin{enumerate}
    \item Computes an Analysis-Hash based on the \texttt{Cobaya} input file and installed code versions.
    \item Signs the hash using the private key.
    \item Stores relevant information in a pre-object JSON file.
    \item Outputs the Analysis-Hash and signature for immediate publication.
\end{enumerate}

The prover should publish the Analysis-Hash, signature, and their public key at this stage.

\subsection{Generating Proof}

After running the analysis chain, the \texttt{proof} command is used to generate the final proof:

\begin{verbatim}
topocobaya proof INPUT.yaml [-p INPUT.yaml]
\end{verbatim}

This command:
\begin{enumerate}
    \item Verifies that the analysis hasn't changed since freezing.
    \item Computes hashes for the employed likelihoods and random seeds.
    \item Processes the output chain to build a Merkle Tree.
    \item Creates and signs a proof object containing all necessary verification data.
    \item Saves the proof object, signatures, and public key to a JSON file.
\end{enumerate}

\subsection{Verification}

The \texttt{verify} command is used by verifiers to check the validity of an analysis:

\begin{verbatim}
topocobaya verify INPUT.yaml [-p INPUT.yaml]
\end{verbatim}

This command:
\begin{enumerate}
    \item Checks the input file against the Analysis-Hash.
    \item Verifies signatures using the prover's public key.
    \item Validates the dataset against the proof object.
    \item If all checks pass, runs the analysis chain.
    \item Verifies the Merkle tree roots in real-time during the analysis.
\end{enumerate}

The verification process continues until a predefined time (set by the user in the \texttt{params.json} has elapsed or a mismatch between the computation and the committed analysis is detected.

\subsection{Workflow}

The typical workflow for a verifiable analysis using \texttt{TOPO-Cobaya} is as follows:

\begin{enumerate}
    \item The prover generates a key pair using \texttt{topocobaya keygen}.
    \item Before running the analysis, the prover freezes the analysis state with \texttt{topocobaya freeze}.
    \item The prover publishes the Analysis-Hash, signature, and public key.
    \item The prover runs the analysis using \texttt{Cobaya}.
    \item After the analysis is complete, the prover generates the proof using \texttt{topocobaya proof}.
    \item The prover publishes the proof object and makes the code and data available.
    \item Verifiers can then use \texttt{topocobaya verify} to independently verify the analysis.
\end{enumerate}

This workflow ensures transparency and reproducibility in cosmological analyses, allowing for independent verification of results while maintaining the integrity of the analysis pipeline.

\section{Application to MCMC Chains}

MCMC chains are excellent candidates for this type of blinding, as the chain itself provides a faithful summary of the code’s status at runtime. The only challenge is that MCMC chains are stochastic; however, this can be resolved by providing fixed random seeds, that can be added to the proof object for verification. 

\subsection{Example Case}

\begin{figure*}
    \centering
    \includegraphics[width=0.66\textwidth]{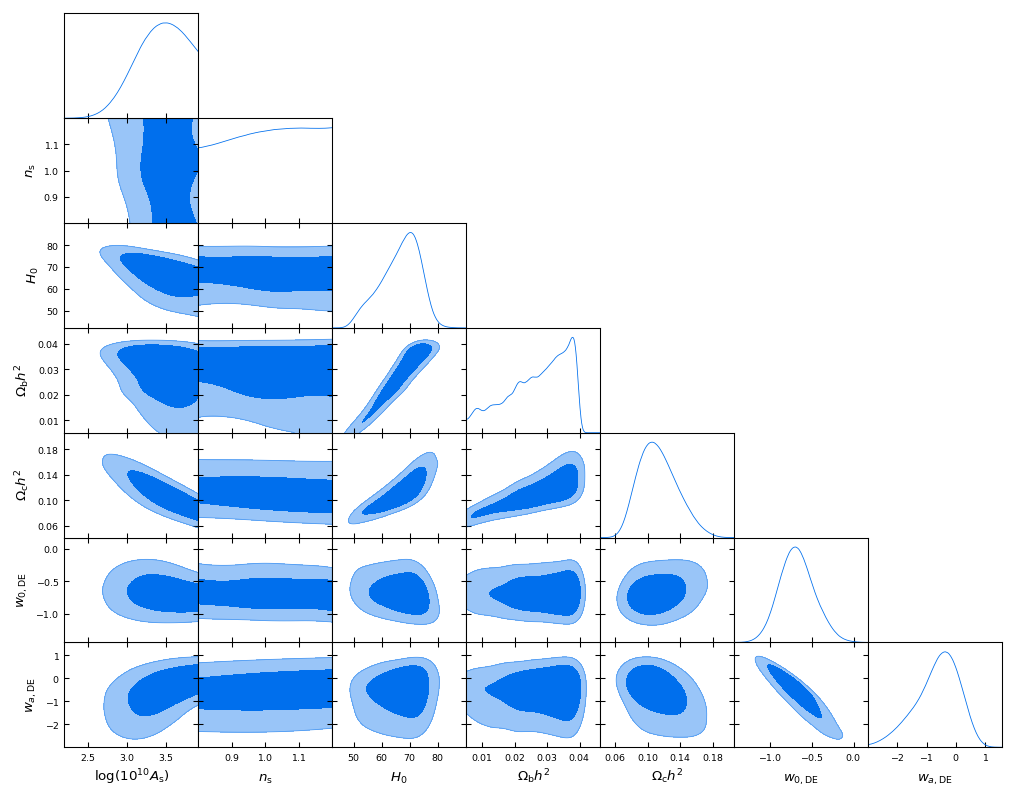}
    \caption{Triangle plot of the posterior distribution obtained from the BAO likelihoods described in the main text. We use \texttt{TOPO-Cobaya}, which essentially has all the functionality of \texttt{Cobaya} plus cryptographic verification.
    The result of this chain can be verified using the proof-object shown in \cref{fig:proofobj}.}
    \label{fig:triangle}
\end{figure*}

In the following, we provide all the data required to validate an MCMC chain that we have run using the input file \texttt{scripts/BAOs.yaml} available in our repository \githubtopocobaya.
This example likelihood uses the code Python wrapper \texttt{classy} of the Einstein-Boltzmann solver \texttt{CLASS} \cite{Blas:2011rf, lesgourgues2011cosmic}.
As an example we run a Baryon Acoustic Oscillations (BAO) cosmological inference analysis using  the following likelihoods:  BAO detection of the 6dF Galaxy Survey \cite{Beutler:2012px}, 
the BAO scale measurement of SDSS DR7 Main Galaxy Sample
\cite{Ross:2014qpa}, the power spectrum LRG+ELG+QSO BAO of SDSS DR16 and the auto-correlation of Lyman-$\alpha$ and the cross-correlation of Lyman-$\alpha$xQSO of SDSS DR16 \cite{Alam:2020sor}.
We vary the following cosmological parameters: $\log(10^{10} A_\mathrm{s})$, $n_\mathrm{s}$, $H_0$, $\Omega_\mathrm{b} h^2$, $w_{0,\mathrm{DE}}$ and $w_{a,\mathrm{DE}}$. We apply the same non-informative priors used by the corresponding collaborations.
The result of this run is shown in \cref{fig:triangle}. Using a Metropolis-Hastings sampler \cite{Lewis:2002ah, Lewis:2013hha} the convergence of this run was achieved after approximately 10 hours on a 16-core Intel-i9 laptop.

We are running the following versions of the codes:
\begin{itemize}
    \item \texttt{CLASS} version: \hash{22b49c0af22458a1d8fdf0dd85b5f0840202551b}
    \item \texttt{TOPO-Cobaya} version: \hash{30c8cf2e6bacf6ef18fb78ba8f15458ff2107611}
    \item Hash of the \texttt{scripts/BAOs.yaml}: \hash{26451e40a8fccce4390699775dcbd3f40c1dd77a0eb865a9e90e19d02997a749}
\end{itemize}
We use a public key (generated with \texttt{topocobaya}) that 
corresponds to the Ethereum address
\hash{0xb6766c8A362a692952Bf6cA6DDc277A99cE49a92}.
This information combined provides the following Analysis-Hash: 
\hash{d4b9d47377300d39c8b1a2c8f1f32ad149aac6fbb093ef9947af47ea69da3ded}. The full pre-commit object, including the full public key and the signature $S_A$ is shown in \cref{fig:precommitobj}. 
Notice that the current git commit hash of our \texttt{TOPO-Cobaya} code can be different due to updates we release on the way \githubtopocobaya. If you want to reproduce these exact numbers, go to the above mentioned commit.

The Analysis-Hash is stored on-chain as a transaction sent from our Ethereum address to the \texttt{TOPO} address \hash{0x0000000000000000000000000000000000007090}
timestamped on Oct-31-2024 12:33:23 PM UTC. This record is publicly accessible on \href{https://etherscan.io/address/0x0000000000000000000000000000000000007090}{Etherscan}.
This approach guarantees that after freezing the pipeline, no modifications can be done in the future. Committing this to a public blockchain, which is essentially irreversible, ensures that no actor can claim any other pipeline was used in the future, since any small modification will alter the Analysis-Hash.
We encourage \texttt{TOPO-Cobaya} users to adopt this approach to ensure all frozen analyses are accessible under a single Ethereum address and can be independently and publicly verified.

Using this frozen analysis, we have ran an MCMC chain and obtained the following proof object shown in \cref{fig:proofobj}. It contains 14 roots of the Merkle Tree, corresponding to a chain of 21281 points, showcasing the effectiveness of Merkle Trees to compress data verification.

\begin{figure*}[ht]
\centering
\begin{verbatim}
    "proof_object": {
        "git_hash": "30c8cf2e6bacf6ef18fb78ba8f15458ff2107611",
        "input_hash": "26451e40a8fccce4390699775dcbd3f40c1dd77a0eb865a9e90e19d02997a749",
        "code_hashes": {
            "classy": "22b49c0af22458a1d8fdf0dd85b5f0840202551b"
        }
    },
    "signatureA": "0x1d8e9a27a4aaaabb33db433bd03f48c309caa0d02542a30b345b749319e26b3a2a
                   2c1b6cff6d70b2aab7e3415295dadb42e4284540461e8ed8fb8d60c6497fc001",
    "signatureB": "None",
    "public_key": "0xb90a9121ef0b683648f056be8fe1993312d057e23d60ce1ee52b6b6e2ff729f374
                   058613fe5bf73859beadfec82736ae23529e2efe65c3eaa988c524244bcac8"
\end{verbatim}
\caption{Example of pre-commit object for freezing the analysis pipeline. It contains the hashes of the used sampler code, of the Einstein-Boltzmann solver and of the input file used to run the likelihood. This is then signed using the public key to produce the \texttt{signatureA} and ensure the pipeline hasn't been tampered with.}
\label{fig:precommitobj}
\end{figure*}

\begin{figure*}[ht]
\centering
\begin{verbatim}
    "proof_object": {
        "roots": [
            "607df5da1108e20a8bbad807551db28e88bc62a7a10d21424e885df67a20adc8",
            "cd273898988a64b8b9705616356e147dc4f9c0d76c2a169b0e210fbf7525d763",
            "aa1e41963916a315f21a0caffded38fc77eefde80efcdcd5f1217b8061432225",
            "8bc93e9506775778e099bbf6b979e88ccfe5d70a83b1805f3b516868c8d46f63",
            "12bb78634ccaa1cfc9d622184c91617f7b9c5a1c7aae3298a22b6fdec5725632",
            "e5f8025174e653b8f1746ccb74e414dcc536aeafd09a7fea7ab1bcf1567c2921",
            "55e4e4f4096309f9c5ec1765e58bb4f2d1ff97b08c9b95901a6fed9aab35fe2d",
            "8f66c856d4ab6f2d310ab9880a714f0f0b3bace956cc22a2b9719d41efbd685d",
            "243df9c9b0247cb87de14b4c29eb26a559d38d7987284ea9b1cb8a59032b36de",
            "e65fcf575d15b61484b8f28336ef75aac1af0d30de989eb1f381804eb259d937",
            "5085b17c733fd979fa63fe704cfc3359f09d2f95e8d7e3913e2575453a98cbc8",
            "fee19fa2e02d34e1bf429588b3780b68e143f8ab440d20aa99adc8eccc8ba549",
            "d041e543e49d4a7c0591f90ab9c879b0465ad841c383711624662f736fdab9f4",
            "ae3b3f85c567140df74d7302514c7caa3c24e326f85ddf11bbc733b22fa892d7"
        ],
        "data_hash": "b3fd43d468762cca571d331cb09905db966eaae050f9e40d394bfb7c4c1de68e",
        "seed": 420,
        "skip": 5,
        "round": 6,
        "Analysis_hash": "d4b9d47377300d39c8b1a2c8f1f32ad149aac6fbb093ef9947af47ea69da3ded"
    },
    "signatureA": "0x1d8e9a27a4aaaabb33db433bd03f48c309caa0d02542a30b345b749319e26b3a2a
                   2c1b6cff6d70b2aab7e3415295dadb42e4284540461e8ed8fb8d60c6497fc001",
    "signatureB": "0x49e496d26c20da090b83f1d65e0ec419fbc29a0cddfa48bc89004ea1e29f904624
                   e5eff80f3e5a1eb1c6a6bdd119824aa16e76f28e13fe3bae462a6f9aaadee501",
    "public_key": "0xb90a9121ef0b683648f056be8fe1993312d057e23d60ce1ee52b6b6e2ff729f374
                   058613fe5bf73859beadfec82736ae23529e2efe65c3eaa988c524244bcac8"
\end{verbatim}
\caption{Example of proof-object for proving honest computation of the MCMC chain. It contains the 14 roots of the Merkle Tree (corresponding to a chain of 21281 points). It also contains the hash of the data and other parameters used in the verification, such as the random seed. Using the same public key as before, the proof-object is signed and the \texttt{signatureB} is generated.}
\label{fig:proofobj}
\end{figure*}

With this information, anyone can now use \texttt{topocobaya verify} to determine whether the chain has been run as claimed, ensuring the Prover is not misrepresenting the results. The novelty of our approach is that the Verifier does not need to run the entire pipeline again (in this case 10 hours), but will progressively see higher levels of the Merkle Tree being verified. After a few levels, the probability that the run has been maliciously manipulated will be infinitesimally low. The user can choose in the \texttt{params.json} for how long (in minutes) the verification run will proceed.

\section{Conclusion}

In this paper, we introduced \texttt{TOPO} (Time-Ordered Provable Outputs), a framework that addresses key challenges in reproducibility, integrity, and trust within astrophysical data analysis. Using deterministic hashing, Merkle Trees, and ECDSA digital signatures, \texttt{TOPO} provides a fully trustless and verifiable method to freeze, execute, and validate data pipelines, all within a transparent and tamper-proof structure.

Our approach ensures that analysis pipelines remain immutable once they claim to be frozen, with cryptographic proofs stored on a publicly accessible blockchain to provide permanent, time-stamped verification. By adopting this methodology, astrophysicists can securely blind and validate their results without relying on external “unblinders” or intermediaries, thus preserving the integrity of the scientific process. Additionally, \texttt{TOPO} and especially its integration with popular likelihood codes with \texttt{TOPO-Cobaya}, can be adopted on an individual basis. This can seamlessly integrate with any blinding schemes already established by research collaborations, improving trust and formal verification.

Through our case study with \texttt{TOPO-Cobaya}, we demonstrated the practical application of this framework in cosmological parameter estimation, illustrating that cryptographically verifiable analyses can be effectively integrated with sophisticated probabilistic methods like Markov-Chain-Monte-Carlo runs.

As scientific findings increasingly influence public policy and technological innovation, we encourage broader adoption of trustless verification methods like \texttt{TOPO} to reinforce both the credibility of scientific research and public trust in scientific findings.

\section*{Credit authorship contribution statement}
\textbf{Santiago Casas}: Original idea, conceptualization, code development, likelihood development, computations and paper writing.
\textbf{Christian Fidler}: Original idea, conceptualization, code development, cryptographic library development,  computations and paper writing.

\section*{Acknowledgements}
We thank the open source community for providing valuable libraries in Web3 that made this code possible. 
SC acknowledges support by the Science and Technology Facilities Council (grant number ST/W001225/1).

\bibliographystyle{aa} 
\typeout{}
\bibliography{refs}


\end{document}